\documentclass[aps, prl, reprint]{revtex4-1}

\usepackage{graphicx, epsfig, array, subfigure, epstopdf}
\usepackage[colorlinks=true, citecolor=blue, linkcolor=blue, urlcolor=blue]{hyperref}
\usepackage[colorinlistoftodos]{todonotes}

\begin{document}

\title{Three-dimensional superconducting resonators at $T < 20$~mK with the photon lifetime up to $\tau=2$ seconds}% Force line breaks with \\
\thanks{This work was supported by the US Department of Energy, Office of High Energy Physics.}%
\author{A. Romanenko} 
\email{aroman@fnal.gov}
\author{R. Pilipenko}
\author{S. Zorzetti}
\author{D. Frolov}
\author{M. Awida}
\author{S. Belomestnykh}
\author{S. Posen}
\author{A. Grassellino}

\affiliation{Fermi National Accelerator Laboratory, Batavia, IL 60510, USA}

\date{\today}

\begin{abstract}
Very high quality factor superconducting radio frequency cavities developed for accelerators can enable fundamental physics searches with orders of magnitude higher sensitivity, as well as offer a path to a 1000-fold increase in the achievable coherence times for cavity-stored quantum states in the 3D circuit QED architecture. Here we report the first measurements of multiple accelerator cavities of $f_0=$1.3, 2.6, 5 GHz resonant frequencies down to temperatures of about 10~mK and field levels down to a few photons, which reveal record high photon lifetimes up to 2 seconds, while also further exposing the role of the two level systems (TLS) in the niobium oxide. We also demonstrate how the TLS contribution can be greatly suppressed by the vacuum heat treatments at 340-450$^\circ$C.
\end{abstract}

%\pacs{}

\maketitle

Superconducting radio frequency (SRF) cavities in particle accelerators routinely achieve~\cite{Padamsee_Ann_Rev_Nucl_2014, Romanenko_APL_2014} very high quality factors $Q > 10^{10}-10^{11}$ corresponding to photon lifetimes $\tau$ as long as tens of seconds. These are much higher than highest $Q\sim10^8$ reported in various quantum regime studies~\cite{Paik_PRL_2011, Reagor_PRB_2016} with $\tau \sim 1$~msec. Thus, adopting SRF cavities for a 3D circuit QED architecture for quantum computing or memory appears to be a very promising approach due to the potential of a thousand-fold increase in the photon lifetime and therefore cavity-stored quantum state coherence times. There is also a variety of proposed fundamental physics experiments, i.e. dark photon and axion searches~\cite{Parker_PRD_2013, Bogorad_PRL_2019, Janish_PRD_2019}, for which the availability of higher $Q$ cavities in the lower photon regime would directly translate into multiple orders of magnitude increases in search sensitivities.

Recent investigations~\cite{Romanenko_PRL_2017} revealed that the two-level systems (TLS) residing inside the niobium oxide may play a significant role in the low field performance of SRF cavities, similarly to the 2D resonators~\cite{Anderson_PhilMag_1972, Martinis_PRL_2005}. For further understanding of the physics involved, and to guide any future $Q$ improvement directions, a direct probing of SRF cavities in the quantum regime is required. Up to now, no such investigations have been performed.

In this article, we report the first measurements of a selection of state-of-the-art SRF cavities down to very low temperatures ($T<20$~mK) and very low fields of a few photons (``quantum'' regime). We achieve the highest reported photon lifetimes of more than 2~sec, and observe a $Q$ decrease when going from previously explored temperatures of 1.4~K down to below 20~mK. It is also the first direct study of the TLS in the 3D Nb resonators in the quantum regime, as well as the demonstration of the drastic TLS-induced dissipation decrease associated with the oxide removal. Our results demonstrate that SRF cavities can serve as the longest coherence platform for e.g. 3D cQED and quantum memory~\cite{Reagor_PRB_2016, Xie_APL_2018} applications, as well as for various fundamental physics experiments, such as dark photon or axion searches~\cite{Parker_PRD_2013, Bogorad_PRL_2019, Janish_PRD_2019}. 

%CAVITIES STUDIED AND THE HIGH TEMPERATURE PERFORMANCE
We have used fine grain high residual resistivity ratio (RRR)$\gtrsim$200 bulk single cell niobium cavities of the TESLA shape~\cite{TESLA_Cavities_PRST_2000} with resonant frequencies $f_0$ of the TM$_{010}$ modes of 1.3, 2.6, and 5.0 GHz. 

\begin{figure*}[htb]
 \includegraphics[width=\textwidth]{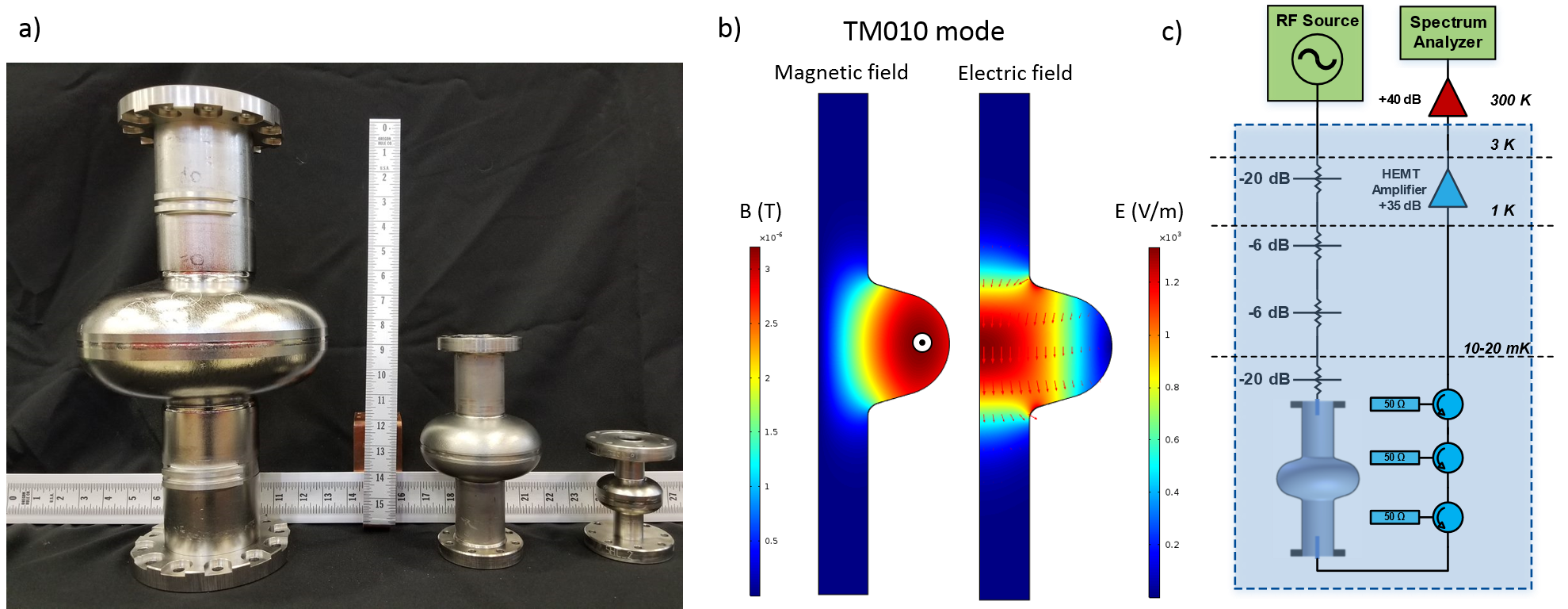}
\caption{\label{fig:Cavities}a) Single cell cavities of TESLA geometry used for the measurements; b) distribution of the magnetic and electric fields in the utilized TM010 mode (half of the rotationally symmetric cavity is shown); the coupling to the mode is performed using the pin couplers on cavity axis on both sides;  c) typical microwave setup used for measurements (the attenuators and the amplifiers were different in some cases for different frequencies and cooldown cycles).}
\end{figure*}

The cavities utilized are shown in FIG.~\ref{fig:Cavities} along with the calculated electric and magnetic field distributions. The fundamental frequency sets the radial dimension $R$ of the cavities: $R \propto 1/f_0$. Electromagnetic coupling to the cavities is performed using axial pin couplers at both ends of the beamtubes. 

We have also applied novel heat treatments in a custom designed furnace~\cite{TLS_patent} to remove the niobium pentoxide (Nb$_2$O$_5$) and to directly investigate the associated improvement in the TLS dissipation on both 1.3~GHz and 5~GHz cavities. The 1.3 GHz cavity has been heat treated at 340$^\circ$C for several hours, whereas the 5~GHz cavity has been treated similarly at 450$^\circ$C as the last step of the cavity preparation.

The measurements have been performed first at the vertical test facility where the cavities are submerged in liquid helium and temperatures down to 1.4~K can be achieved, and then in the dilution refrigerator at temperatures down to 10-20~mK. 

\begin{figure}[htb]
 \includegraphics[width=\linewidth]{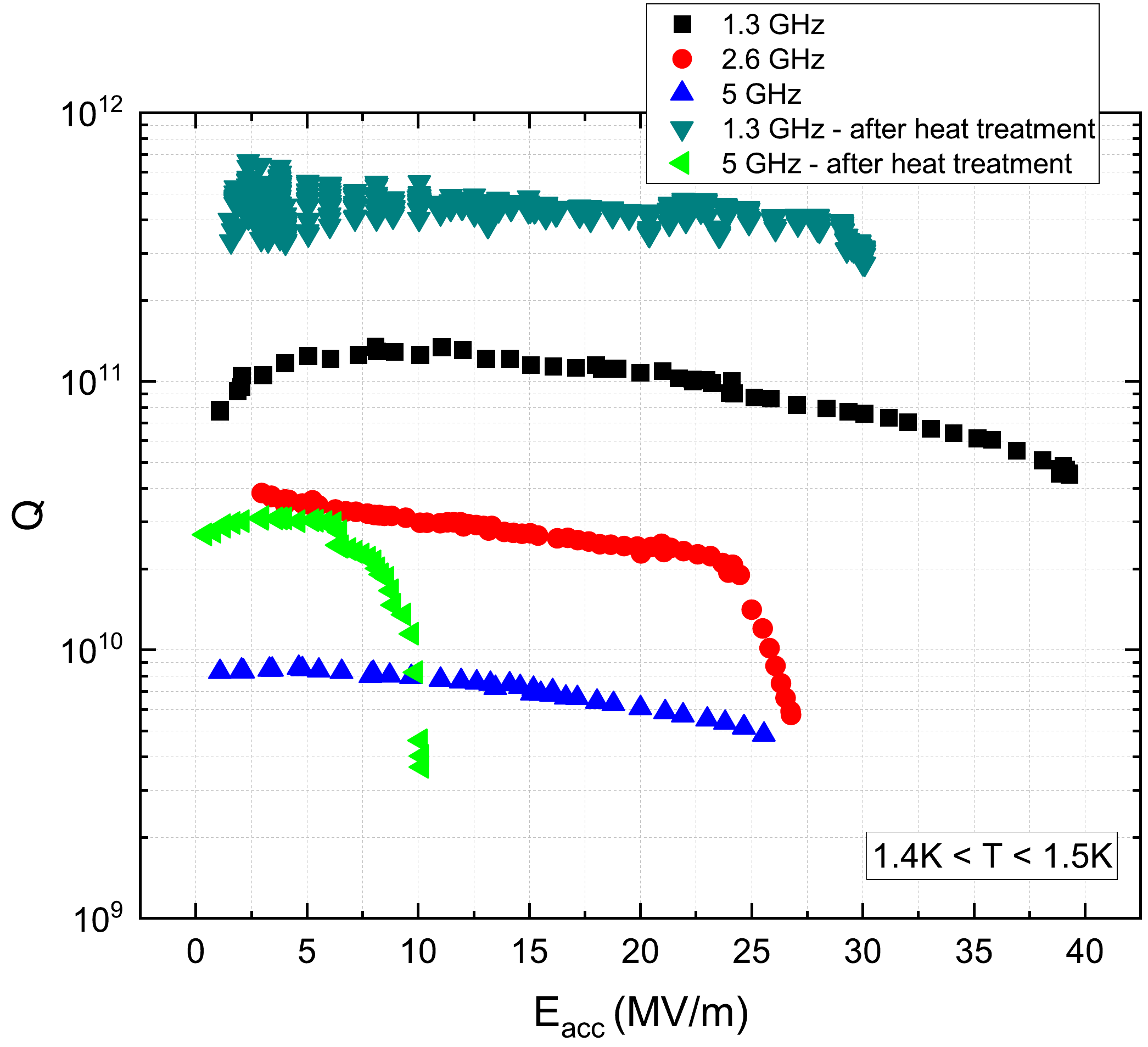}
\caption{\label{fig:Q_E}Intrinsic cavity quality factors of the investigated 1.3, 2.6, and 5~GHz cavities as a function of the accelerating field at temperatures $1.4<T<1.5$~K.}
\end{figure}

For cavities in the vertical test dewar we have used the standard SRF measurement techniques~\cite{Melnychuk_RSI_2014} at higher accelerating fields $E$, and the filtered decay method~\cite{Romanenko_PRL_2017} at lower fields. The $Q(E)$ results at temperatures down to 1.4~K in a broad range of - higher - cavity fields are shown in FIG.~\ref{fig:Q_E}. It is remarkable that the 1.3 GHz cavity after the 340$^\circ$C heat treatment has an extremely high quality factor $Q \gtrsim 4 \times 10^{11}$ in the broad range of fields, higher than the previously reported highest results~\cite{Romanenko_APL_2014}. This indicates that the 340$^\circ$C heat treatment suppresses the residual resistance at all fields, which may be also related to the TLS or other potential mechanisms, i.e. to the elimination of the possible metallic niobium suboxide inclusions inside the pentoxide layer.

For dilution refrigerator measurements, the double layer magnetic shielding around the full cryostat was used and the magnetometers placed directly on the outside cavity surfaces indicated that the DC ambient magnetic field level was shielded to below 2~mG in all cases. The microwave setup included a series of attenuators on the cavity input line, as well as both cryogenic and room temperature amplifiers on the pickup line. For one of the runs with the 5 GHz cavity the Josephson Parametric Amplifier (JPA) has been used in the output line as well. The measurement configuration including the low-noise cryogenic amplifier allows to measure reliably the photon lifetimes down to the average cavity population $\overline{n}\sim$10 photons, while JPA enabled extending it further to single photon levels. The cavity placement and a typical microwave schematic of the setup are shown in FIG.\ref{fig:Cavities}. 

The average photon number is calculated from the cavity stored energy $U$: $\overline{n}=U/\hbar\omega$, where $U=P_\mathrm{t} Q_\mathrm{t}/\omega$ is extracted from the measured transmitted signal $P_\mathrm{t}$ at the pickup coupler with the external quality factor $Q_\mathrm{t}$.

\begin{figure}[htb]
 \includegraphics[width=\linewidth]{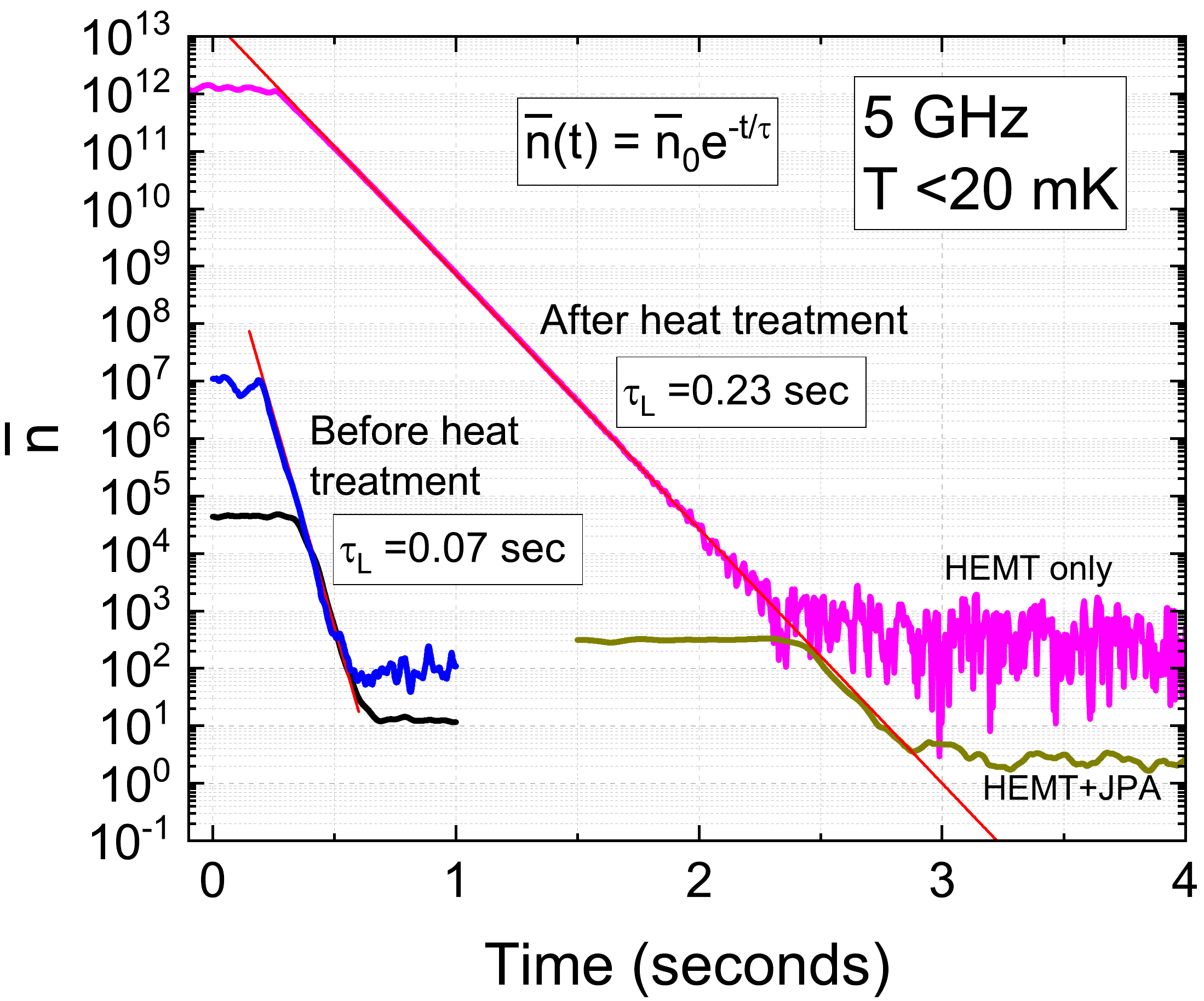}
\caption{\label{fig:N_t}Average photon population decay upon switching the RF power off measured in the 5~GHz cavities before and after heat treatment at 450$^\circ$C. Blue and black curves correspond to the decays from different starting fields and with the different filtering bandwidths (100 Hz and 10 Hz respectively). The red lines show linear fits for both. Magenta and dark yellow lines show stored energy decays of the 450$^\circ$C treated cavity with the much longer photon lifetime.}
\end{figure}

\begin{figure}[htb]
 \includegraphics[width=\linewidth]{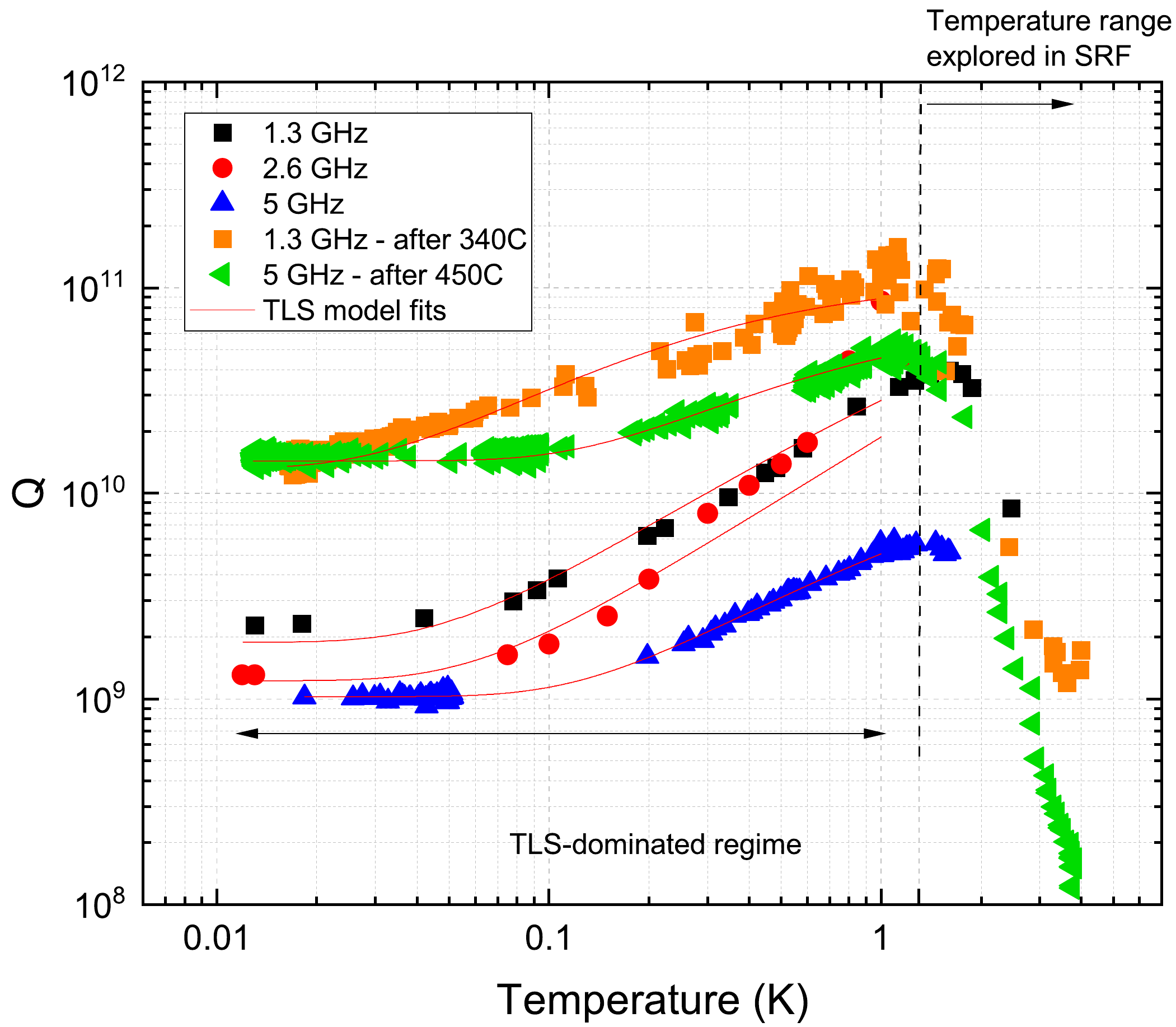}
\caption{\label{fig:Q_T}Intrinsic cavity quality factor $Q$ of the investigated 1.3, 2.6, and 5~GHz cavities as a function of temperature. Previously, SRF cavities have only been studied at temperatures above about 1.3~K. Below about 1~K a significant decrease in $Q$ is observed consistent with the TLS dissipation - the red lines show TLS model fits. A dramatic increase in $Q$ associated with the oxide-modifying heat treatment is apparent on the 1.3 GHz and 5 GHz cavities.}
\end{figure}

The typical decay curves for $5$~GHz cavities before and after 450$^\circ$C vacuum heat treatment for different starting cavity photon populations are shown in FIG.~\ref{fig:N_t}. Blue and black curves correspond to two different starting power levels, as well as the different resolution bandwidths of the spectrum analyzer. As the exponential decay fits (red lines) indicate, the time constant and therefore the quality factor $Q$ remains constant down to the noise floor of about 10 photons and $\sim$2 photons. We have also observed no rf field amplitude dependence of the $Q$ factor for all the cavities in the dilution refrigerator setup. It is consistent with our previous studies and higher temperature/higher field measurements in the current study, which showed that the ``critical'' TLS saturation field $E_\mathrm{c}$ for niobium oxide is much higher - of the order of $E_\mathrm{c} \sim 0.1$~MV/m  - and therefore TLS are not saturated by the microwave fields from about $\overline{n}\sim10^{20}$ all the way down to $\overline{n}\sim2$.  

$Q(T)$ measurements, which represent the main findings of our paper, are shown in FIG.~\ref{fig:Q_T}. A characteristic $\propto 1 / \tanh \left(\alpha \frac{\hbar \omega}{2 k T}\right)$ temperature dependence of the quality factors $Q(T)$ for all the cavities is clearly observed with the $Q$ decreasing towards lower temperatures. The amount of $Q$ degradation is drastically suppressed by the heat treatments - 340$^\circ$C for the 1.3~GHz and 450$^\circ$C for the 5~GHz cavity respectively. This is consistent with the removal of the significant number of TLS, which are hosted by the pentoxide layer of SRF cavities - as shown in our previous work~\cite{Romanenko_PRL_2017}. 

Below about 1~K the contribution to the surface resistance caused by the thermally excited quasiparticles becomes negligible and the $Q(T)$ curves appear to be dominated by a dissipation caused by the TLS. The TLS dissipation increases as the temperature is further lowered due to the decreased thermal saturation and therefore an increased number of the TLS systems participating in the resonant absorption of the microwave power.

In the TLS-dominated regime, an excellent fit is obtained using the ``standard'' TLS model~\cite{Anderson_PhilMag_1972, Martinis_PRL_2005} dissipation with $\delta_0$ as the loss tangent of the TLS at $T=0$~K, an additional coefficient $\alpha$ to account for temperature measurement efficicency, and a fixed residual ($R_\mathrm{res}$) surface resistance:
\begin{equation}
\frac{1}{Q(T)} = F \delta_0 \tanh \left(\alpha \frac{\hbar \omega}{2 k T}\right) + \frac{R_\mathrm{res}}{G}
\end{equation}
where $F$ is the calculated filling factor~\cite{Reagor_PRB_2016, Gao_APL_2008, Wang_PRL_2009}, $G = 268$~Ohm is the geometry factor of the $TM_{010}$ mode for TESLA shape extracted from the finite element simulations~\cite{TESLA_Cavities_PRST_2000}. 

In Table~\ref{Fit_summary} the obtained fit $F\delta_0$ values are shown for all the cavities investigated. A dramatic - about an order of magnitude - decrease in $F\delta_0$ is associated with the heat treatments. For cavities before the heat treatments (with the pentoxide layer) the participation ratios can be calculated as in Ref.~\cite{Romanenko_PRL_2017} assuming the $\sim$5~nm oxide layer, and the $\delta_0$ values can then be estimated as well. These are listed in Table~\ref{Fit_summary} wherever applicable.

%TABLE OF FIT RESULTS
\begin{table}
\caption{\label{Fit_summary}Summary of TLS model fitting results.}
\begin{ruledtabular}
\begin{tabular}{*5c}
$f_0$ (GHz) & Oxide treatment & $F \delta_0$ & $F$ & $\delta_0$ \\
\hline
1.3 & No & $5.2\times10^{-10}$ & $1.0\times10^{-7}$ & 0.17 \\ 
2.6 & No & $8.2\times10^{-10}$ & $2.4\times10^{-8}$ & 0.13 \\
5 & No & $9.1\times10^{-10}$ & $1.2\times10^{-8}$ & 0.08 \\
1.3 & 340$^{\circ}$C 3 hrs & $6.7\times10^{-11}$ & unknown & n/a\\
5 & 450$^{\circ}$C 3 hrs & $5.6\times10^{-11}$ & unknown & n/a \\
\end{tabular}
\end{ruledtabular}
\end{table}%

While the presence of TLS leads to the decrease of $Q$ from its values at higher temperatures, even in the worst case (5~GHz without heat treatment) we obtain the photon lifetime $\tau=32$~msec, which is several times higher than the previous record of $\sim$10.2~msec~\cite{Reagor_APL_2013}. After the heat treatments, the achieved photon lifetimes of 0.5-2~sec correspond to a $\sim$50-200 times improvement.

%SIMS studies

\begin{figure}[htb]
 \includegraphics[width=\linewidth]{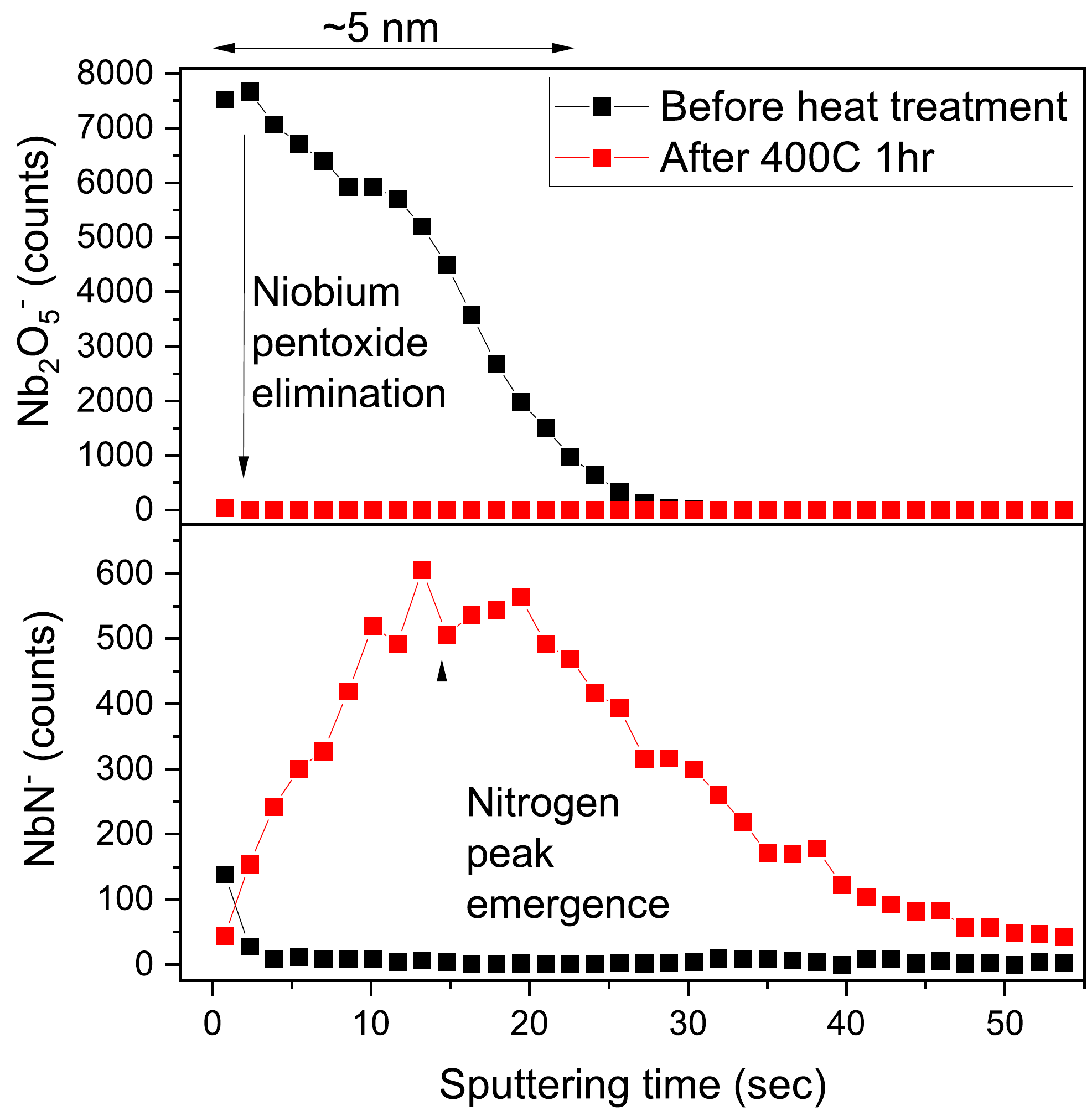}
\caption{\label{fig:SIMS}TOF-SIMS depth profiles obtained on the cavity cutout revealing the changes in the niobium pentoxide and nitrogen-related signals after 400$^\circ$C in situ heat treatment. About 5~nm thick Nb2O5 pentoxide layer is completely dissolved after the 400$^\circ$C treatment. Another apparent change is the increase in the nitrogen level within $\sim$10~nm from the surface.}
\end{figure}

To reveal the underlying material changes happening during the vacuum treatment in the temperature range of interest (340-450$^\circ$C) we have performed the direct studies on the niobium cavity cutout using the in-house time-of-flight secondary ion mass spectrometry (TOF-SIMS) system. TOF-SIMS allows obtaining the depth profiles of various elements within the sample with the sub-nanometer depth and better than parts-per-million concentration resolution and has been actively used in recent years to guide the SRF cavity near-surface structure tailoring~\cite{Romanenko_SRF2019_THP014}. Shown in FIG.~\ref{fig:SIMS}, the comparison before and after the 400$^\circ$C vacuum treatment (without subsequent air exposure) has confirmed the removal of the Nb$_2$O$_5$, likely explaining the reduced TLS dissipation after 340-450$^\circ$C treatments. Furthermore, we have discovered the emergence of the strong near-surface nitrogen enrichment, which is the likely cause of the ``doping''-like effect we have found at higher cavity fields after these heat treatments~\cite{Posen_midT_2019_arXiv}.

It is intriguing that the cavities after 340-450$^\circ$C heat treatments still have some non-zero $Q$ degradation at temperatures lower than $1$~K (FIG.~\ref{fig:Q_T}). Since SIMS studies suggest that there is no Nb$_2$O$_5$ after these treatments, an additional source of TLS should be present as well. Some potential examples could be other types of niobium oxides (e.g. NbO) and their interfaces with the underlying bulk, or e.g. surface adsorbates. Pinpointing these remaining sources would be a key goal of the future detailed investigations.

Practically, our findings open up a pathway of exploring coupled SRF cavity-transmon structures as the highest coherence superconducting quantum circuits for quantum computing. In particular, implementing the protocol from Ref.~\cite{Heeres_PRL_2015} would allow direct generation of very long-lived Fock states in SRF cavities. One important question is what is the best way to insert the transmon in the SRF cavity to provide enough coupling to the cavity mode of interest while not degrading the ultra-high $Q$. A possible solution is using the low-loss dielectric rod (such as sapphire) to hold the transmon in the relevant field area of the cavity. Corresponding electromagnetic design work and the mechanical and microwave measurements to validate this concept are currently on the way.

For new physics searches, using SRF cavities with the $Q$ factors we have demonstrated allows for a sensitivity increase of multiple orders of magnitude. A prototype dark photon ``light-shining-through-the-wall'' search experiment of the type as in Ref.~\cite{Parker_PRD_2013} but now with much higher $Q$ SRF cavities has been assembled and is currently being commissioned with the results to be reported in future publications. 

In summary, we have performed the first measurements of the state-of-the-art SRF accelerator cavities in the quantum regime and have demonstrated the photon lifetimes as high as $\tau=2$~sec - about a factor of 200 higher than the records so far in this regime. We have also revealed the quality factor decrease at lower temperatures, consistent with the contribution of the TLS hosted by the niobium oxide, and demonstrated its mitigation by the in situ heat treatments at 340-450$^\circ$C resulting in the removal of niobium pentoxide, as witnessed by TOF-SIMS.

\acknowledgments
Fermilab is operated by Fermi Research Alliance, LLC under Contract No. DE-AC02-07CH11359 with the United States Department of Energy. The authors would like to acknowledge Dmitri Sergatskov, Oleksandr Melnychuk, and Damon Bice for participation in some of the aspects of this work, Joe Lykken for his support of this work, as well as David Pappas, Haozhi Wang, and Mustafa Bal for providing a JPA used for one of the measurements.

%\bibliography{../../BIBLIOGRAPHY/ROMANENKO_bibliography}

\end{document}